# VCS: Value Chains Simulator, a Tool for Value Analysis of Manufacturing Enterprise Processes (A Value-Based Decision Support Tool)

Magali Bosch-Mauchand · Ali Siadat · Nicolas Perry - Alain Bernard

Mechanical Systems Engineering, UTC, Compiègne, France
e-mail: magali.bosch@utc.fr
A. Siadat
LCFC, CER Metz - Arts et Métiers Paris Tech, Metz, France
N. Perry
LGM2B/IUT GMP, Université Bordeaux 1, Gradignan, France
A. Bernard
IVGI Team, IRCCyN, Centrale Nantes, Nantes, France

Abstract:

Manufacturing enterprises are facing a competitive challenge. This paper proposes the use of a value chain based approach to support the modelling and simulation of manufacturing enterprise processes. The aim is to help experts to make relevant decisions on product design and/or product manufacturing process planning. This decision tool is based on the value chain modelling, by considering the product requirements. In order to evaluate several performance indicators, a simulation of various potential value chains adapted to market demand was conducted through a Value Chains Simulator (VCS). A discrete event simulator is used to perform the simulation of these scenarios and to evaluate the value as a global performance criterion (balancing cost, quality, delivery time, services, etc.). An Analytical Hierarchy Process (AHP) module supports the analysis process. The value chain model is based on activities and uses the concepts of resource consumption, while integrating the benefiting entities view point. A case study in the microelectronic field is carried out to corroborate the validity of the proposed VCS.

## Introduction

Competition between firms in a same market and field is increasing all the time. Therefore, all efforts need to concentrate on how to obtain a competitive advantage. Great efforts must be done to improve enterprise organisation in order to ensure the durability and prosperity of the enterprise.

In terms of competitive advantage, the value chain approach developed by Porter proposes a vision by activities [1]. The adopted point of view concentrates on strategic decisions. Therefore, the weak point is that this position cannot be considered in the design phase.

To be competitive, the industrial organisation needs to reconsider its own activity and to increase its activity result. Continuity and growing challenges are the aims of all industries. Different ways to reach this goal can be followed.

Berrah et al. said that the concept of performance is aimed at durability but with different requirements [2]. On the one hand, to achieve this aim the enterprise needs to be considered as a whole. On the other hand, it is necessary to make permanent and continuous improvements of the enterprise with regard to its structure and its means, as well as its organisation, management, flows, etc.

Focusing on the design of a decision support tool, our research work concentrates on performance analysis in an industrial context, based on value concepts and models [3-5]. The evaluation process is crucial when we want to carry out a performance analysis that is the source of decision processes based on relevant criteria.

Cost is one of the major performance criteria taking into account by benefiting entities. Moreover, in the design phase, this criterion is a growing concern since cost reduction is crucial for enterprise durability. In order to use a global indicator, this cost point of view will not be the only criterion to be taken into account [6]. Value becomes a common concept, integration more focuses than only cost. That's why the value indicator has to be considered: it aggregates cost, quality, time to delivery, and function satisfaction criteria. So, value is defined as a balanced sum of criteria. But depending to the point of view, the value evaluation of a product or a service can be highly different. It results a need of managing the multi view point results.

## Problem statement

*Why Value? And what is its definition?*

Our purpose is to make relevant decisions on product design and manufacturing processes using a tool-based method of value estimation for each benefiting entity (id. es. stakeholders). In this context, a benefiting entity is defined as the entity or the human organisation that takes advantage of the value created by the enterprise and/or that can have an impact on this value. There is a need to react at the beginning of the product development process in order to make early relevant decisions.

In the design and manufacturing planning phases, we try to enrich the Value Engineering method using an integrated method and different tools to complete the value-based decision support.

Financial data is no longer the discriminatory argument in decision and negotiation. For instance, product and service quality (regarding delivery time) ensure client loyalty, which is a strong point in a competitive context. It is crucial to understand that cost performance is not the only criterion to be taken into account in order to define the best technical product solution and the best associated-manufacturing process. Some factors such as quality and delay create complementary constraints on performances. Therefore, performance indicators (relative to multiple objectives) need to be measured.

Moreover, the value criterion depends on the point of view of the benefiting entities. Each of them has its own judgment of the major criteria for value evaluation.

The use of discrete event simulation contributes to performance evaluation by adding contextual performances. The need to know the results of an enterprise in terms of its activity performance and its technical and managerial processes performance is increasing due to competitive markets. Furthermore, the impacts of decisions must be forecast and evaluated. An analysis of the manufacturing performances must be done to underline the best decisions to make such as to create the best manufactured product value. But the performance cannot be evaluated easily. The potential performance of a manufacturing system depends on many criteria.

A well-argued decision process is strategic when considering the definition of manufacturing process plans and technological development in the production systems of the company [3; 7]. It becomes crucial to have a global performance evaluation. The ISO 14000 constraints emphasize the necessity to proceed with this global evaluation when taking efficient decisions. Hence, the simulation process will help to compare different solutions.

Depending on the point of view of the benefiting entity (client, manufacturer, supplier, designer, commercial engineer, etc.), the performance is evaluated in different ways and by different criteria. It is necessary to work on complementary aspects for instance cost / price / value [9] in order to compare all points of view. In this work, value and its different understandings are crucial in order to evaluate the enterprise activity. So, a hierarchy of the components of the value criterion is proposed for this evaluation analysis.

***The main issue of our decision process based on value criterion is: how to support a value- based simulation approach in the product design phase?***

In the following sections, the research results on Value Chain modelling, simulation and evaluation are depicted [3]. The tools and methods system, called Value Chains Simulator (VCS) is presented in two phases, and its use is described and explained by a case study.

The first section of the paper focuses on related research on performance indicators, on value chains and value concepts, on simulations for organisation improvements (such as lean manufacturing) and on a Multi-Criteria Decision Making (MCDM) approach (the AHP process). This review will highlight the already existing concepts and methods used to perform value simulation and analysis. Then we shall point out the lacks of information or the misunderstanding that should be improved on in order to propose a global value model and a value chain simulator.

The second section deals with the architecture of the proposed tool, the Value Chains Simulator (VCS) and the description of the method used to estimate values. The methodology for using VCS is depicted by, on the one hand, the formalism designed to facilitate users' communication and modelling phases, and on the other hand, the formalization of the structure and deployment of the tool. To carry out the evaluation, the AHP module configuration is introduced in details. To end the VCS description, the evaluation process and models are explained.

The last section relates to the application and illustrates the results of the VCS approach in a case study in the microelectronics field.

## Related Research

### Evaluation process: Performance indicators definition

The definition of the performance indicators of a system is contingent on this system. They are dependent on each other. A performance indicator system is contextual.

The performance evaluation of manufacturing processes is generally performed by using models developed from a particular point of view (temporal, financial, etc.). Bennour and Crestani postulate that human competences are the source of value creation [10]. Therefore, they link human capacities and process performances (in terms of quality, time or financial performances).

For manufacturing enterprises, a classification of the most useful performance criteria has been established in order to determine what criteria must be evaluated to give a measure of the product value and also the value provided by the value chain.

The product value is seen as a global indicator that relates to functional, structural and behavioural aspects [8] [11].

The value of an enterprise object is relative to the benefiting entities (for instance, the shareholders). The benefiting entities can benefit or act on this value. They can be external and/or internal actors of the enterprise. Their appreciation of the product and enterprise performance is modelled in order to give a value indication. The appraisal of value is the result of a combination of several performance indicators such as cost, quality and time [12]. These criteria could be mixed with more subjective indicators such as customer perception, associated services, and environmental impact.

Performance and value indicators, presented in Figure 1, come from a reflection on the benefits of product manufacture for each benefiting entity [4].

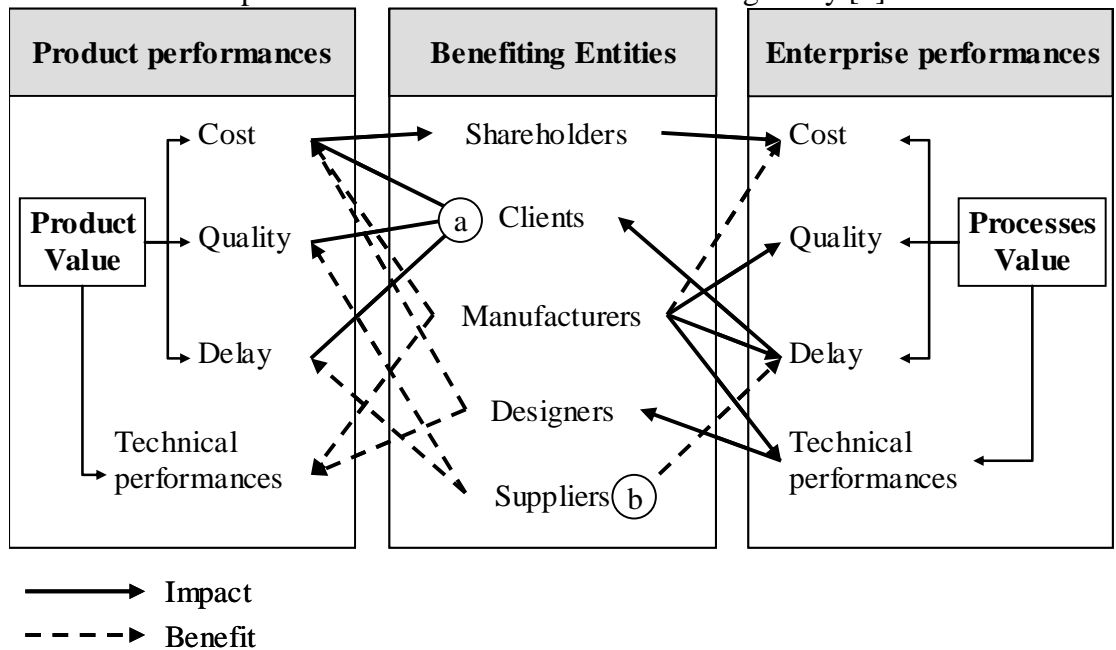

Figure 1. Performances that impact value and their interactions with benefiting entities.

From the client point of view, technical performances are required and must be respected. The cost, quality and delivery time can evolve and so the client value (element "a" in Figure 1) of a product can increase or decrease. From the point of view of the supplier (element "b" in Figure 1), the process value benefits from his delivery time and product quality since it can create manufacturing delays or increase the scrap rate of products.

A link and a parallel can be established between some kinds of performance such as cost and value. To explain this idea, Perrin describes a product by its two distinct views (left and centre blocks of Figure 2) [13].

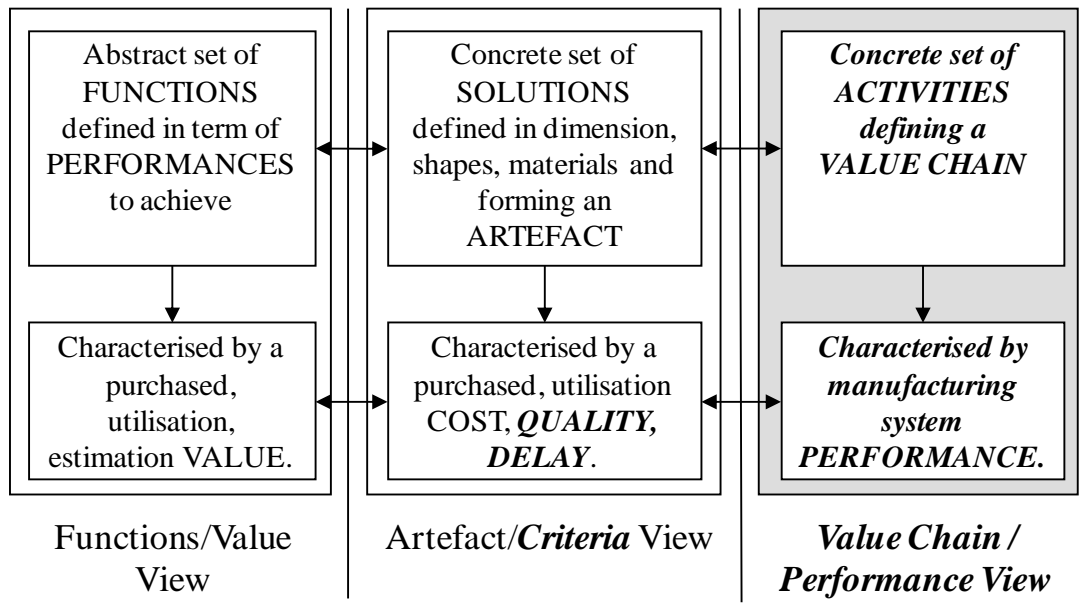

Figure 2. The three views of a product

- The first view, functions/value, depicts the product appreciation in terms of utility. Product value characterizes the expected product performance relative to the required functions.
- The second, artefact/cost, describes the physical object produced to respond to this utility [14]. The relation between the solutions and the activities that could be applied to realise them is not underlined. In order to evaluate the product, a set of activities is defined to characterize the realization of the technical solutions that compose a product.

So our proposal is to add a third face Value Chain / Performance that gives the concepts to be able to link the product solution to its development process and to add some criteria to the second view [5].

Consequently, we propose to add to the Perrin proposal a third view to perform a pertinent evaluation of product functions. This provides the elements needed to evaluate the product value of the manufacturing system performance. A set of activities is defined to characterize the realization of the technical solutions that make up a product.

**Evaluation process: Value chain and value notions integration**

Value is a multi point of view and multi-criteria concept. Elhamdi underlines this aspect of multi points of view by introducing the notion of benefiting entities [15]. In the context of Value Analysis, AFNOR X50-151 standard gives the definition of the term "value" as: "The value is the judgement related to the product on the basis of the user's expectations and motivations, expressed by a ratio which increases when, all other things being equal, the satisfaction of the user's need increases and/or the expenditure related to the product decreases"[1] [14].

---

[1] Translation of « jugement porté sur le produit par l'utilisateur sur la base de ses attentes ou de ses motivations. Plus spécialement, grandeur qui croît lorsque la satisfaction de l'utilisateur augmente ou que la dépense afférente au produit diminue. »

"In competitive terms, value is the amount buyers are willing to pay for what a firm provides. Value is measured by total revenue, a reflection of the price a firm's product orders and the units it can sell" [1] p. 38.

To define "value", we propose the following, based on the definition by Lonchampt [16]:

*Value reflects the judgement related to the product on the basis of the user's expectations and motivations, expressed by a ratio which increases when, all other things being equal, the satisfaction of the user's need increases and/or **the consumption of resources required by all the life cycle phases decreases***.

The characterization of the processes that create value is strategic in order to estimate the impact of a given production on the triptych Value/Cost/Risk. An approach based on value chain modelling and simulation is used to support the manufacturing processes performance analysis. This notion of value is essential for analyzing and comparing the value chain and its components with the same criterion.

Value nets can be used to model the enterprise dynamic regarding material and information flow [17]. Their purpose is to evaluate various alternatives driven by high-level decision criteria to assess the managerial decisions relative to these criteria.

There is a need to structure the concepts in order to create a basic model for value chains that could support performance criteria evaluation and thus, value assessment.

Since the modelling should be used in the design phase, the industrial system is modelled by a Process / Product / Resources model coupled with a Function / Behaviour / Structure approach [11]. This enables us to describe the product and a part of its relative life cycle phases.

The proposed approach based on value chain modelling and simulation is used to support the manufacturing processes performance analysis.

From our point of view, the value chain is a tool of analysis and evaluation that takes into account all the activities that create value. We model these chains by a breaking down of the industrial system into relevant activities regarding strategic, organisational and operational plans. The purpose is to understand the variation of performance with product characteristics and with possible manufacturing process choices (constrained by these characteristics). A simulation of the activities effects is carried out based on the modelling from the supplying of raw materials or parts to the delivery of the final products, as well as the manufacturing process plan. The decisions have an impact on the value chain. These decisions change the performance level of the enterprise.

For the benefiting entities, specific evaluation criteria are defined. For instance, the information needed by the commercial engineer is relevant to global indicators such as manufacturing product cost and delivery time; the designer is much more interested in functions or cost of technical solutions; the manufacturing expert looks at manufacturing activity and process cost; the cost analyst must obtain analytical information.

The proposal of a value chain model is given in order to create the support of the manufacturing performance modelling and evaluation of activities. The first choice of such a model is justified and explained with regards to the related research and to our industrial experience.

The understanding of the value concept is a pre-requisite for using the value chain approach. As said previously, the appraisal of value is the result of a combination of several performance criteria: cost, quality, time and functional satisfaction.

Cost is a problematic value criterion. The relevance of its estimate is a critical point. For this reason, cost estimation is explained in a following section.

## Model and simulation: Simulation tool for lean manufacturing

It is stated that the use of a discrete event simulation tool can permit to reduce costs and also to provide a training tool [18]. In this lean manufacturing context and in this intra-enterprise context, this tool can analyse the effects of the introduction of lean manufacturing improvements into a particular industrial context. The authors' proposal is based on a value chain mapping simulation generator. They analyse Value Stream Mapping constructs symbolised by standard icons used to describe only manufacturing processes. A new classification is highlighted, the Value Stream Mapping Paradigm (VSMP). Then, each icon (graphical representation) is linked to a specific module that models the function and behaviour of the atom (manufacturing process modelling entity).

In an inter-enterprise context, Rabelo et al. propose a very original approach [19]. They combined discrete-event simulation with a dynamic system to model the service and manufacturing activities of the global supply chain of a multinational construction equipment corporation. Moreover, the AHP process is integrated to overcome the potential limitations inherent in simulation models and to take into account qualitative criteria in the decision making process of the managers. Discrete-event simulation is applied to model the manufacturing function and the operational level tasks.

## Results analysis: AHP process

Few Multi-Criteria Decision Making (MCDM) approaches have been developed to deal with complex decision problems. The AHP [20] [21], the Multiple Attribute Utility Theory (MAUT) [22] and Out-ranking Methods [23] are well-known and typical of multiple criteria decision making methods.

The AHP method was developed by Saaty (1981) and it consists of a systematic approach based on breaking down the decision problem into a hierarchy of interrelated elements [20]. This method enables us to take advantage of experts' experience and expertise. The formalization uses a tree of criteria that is established to compare several solutions. At each node of the tree, there is a decision matrix where the pair-wise comparison is entered.

In our module, the results of the simulation provide the value of the selection attributes on which the comparison is based. These results are compared and their scores are evaluated in the scaling system. The expert needs to mark the priority system. In Table 1, the nine-scale comparative points are explained.

| | |
|---|---|
| 1 | Equal importance for both elements compared |
| 3 | Element A is moderately important compared with Element B |
| 5 | Element A is strongly important compared with Element B |
| 7 | Element A has demonstrated importance compared with Element B |
| 9 | Element A is extremely important compared with Element B |

Table 1. AHP scale.

In the AHP process, the deficiency of a criterion can be compensated by the good performance of another criterion that characterizes an alternative. So the process gives a compromise between various alternatives among all their attributes. Then, this process is adapted to value evaluation.

Specific performance indicators have been chosen to evaluate the value of each benefiting entity by using a modified AHP method. These values give the basis of the argument and the negotiation in order to make relevant decision on product functions.

The following section describes the method, the concepts and their structuring and the application tool which has been developed.

# Tool Architecture

### Method proposal and limitations

The proposed value evaluation integrates only measurable and observable data (i.e. tangible value, like production volume or scrap rate for quality evaluation) and not inaccurate and subjective aspects (such as perceptive or aesthetic values). Moreover, the scope of the proposed approach has been limited to the manufacturing phases of the product life-cycle. The first steps (pre-design, detailed design etc.) cannot be modelled and integrated since they cannot be evaluated by the expected value indicator measured in this work.

The proposed system is based on the evaluation of product modelling and of the performance criteria. All these processes are supported by an application tool (Figure 3).

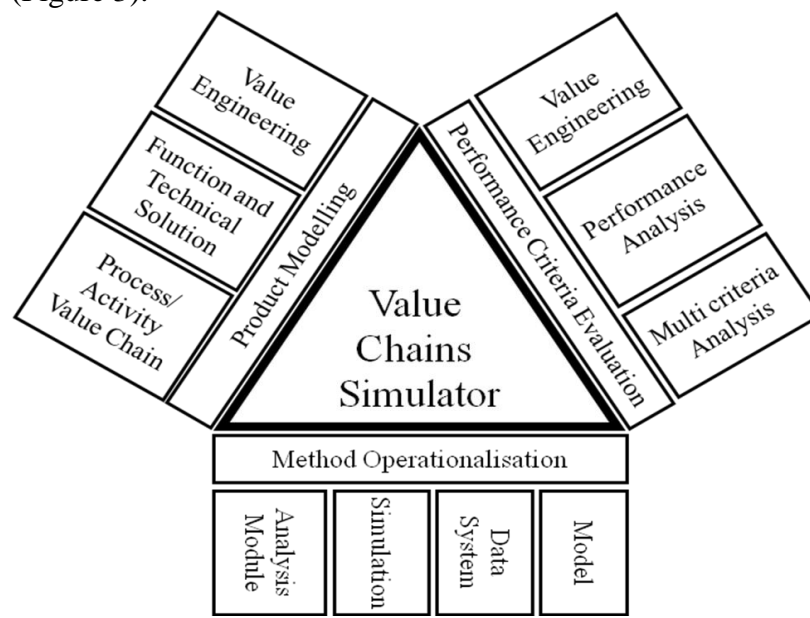

Figure 3. Value Chains Simulator Architecture

Product modelling combines the requirement of the functions and technical solutions specifications and the requirement of the value chain alternatives defined by activities and processes. The Value Engineering process provides the descriptions of the product and the performance evaluation process is based on these descriptions. This gives indicators for Value Engineering decisions process by using performance analysis and multi criteria analysis. In terms of the application tool (see Figure 4), the description of the product is registered in a database.

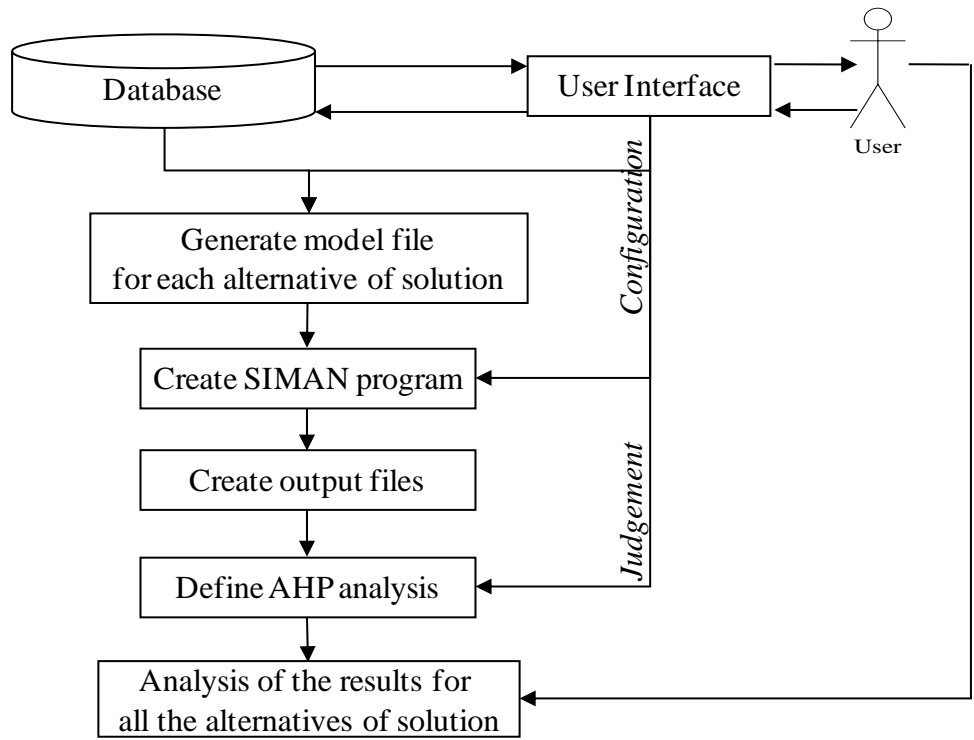

Figure 4. Integration of the simulation and the AHP process in the decision process.

The simulation module gives the results exploited by the analysis module (AHP analysis). This result supports the performance evaluation of each value chain.

The AHP method was chosen for its previous applications for design functions and manufacturing processes selection and for its qualitative and quantitative attributes. Other MCDM methods can gives good results such as Choquet integrals.

## Concepts involved in the modelling

The goal is to evaluate the manufacturing system performance regarding the creation of value that defines the value chain. The concepts involved in this evaluation are as follows:

- **Industrial system modelling**: value chain, activities, resources, expert knowledge, industrial potential;
- **Product modelling**: product, functions, components;
- **Performance criteria**: value, cost, time, quality, reference (objectives), functional satisfaction;
- **Point of view** (benefiting entities): shareholder, client, supplier, manufacturer, designer, etc.

The product is manufactured using resources in a set of activities (the process). The process generates a sum of value (i.e. the value chain). Finally, this value chain makes the industrial potential operate in order to respond to the expected functionalities and the industrial potential limits the product. In fact, the production process that meets the product requirements varies depending on the industrial potential requested.

The product must give solutions to the functions required by the functional specifications. These functions consume resources such as components (material resources) to realize the technical solutions (Figure 5).

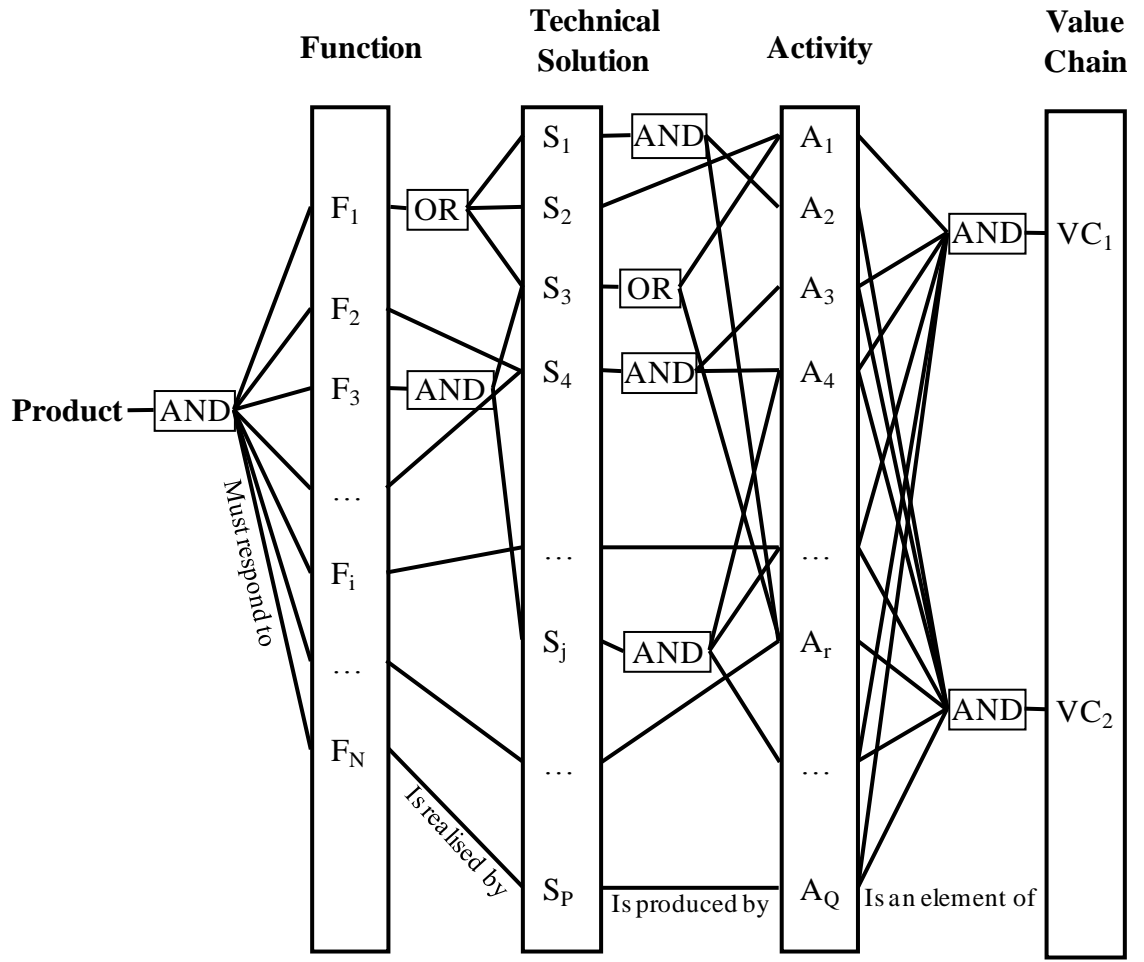

Figure 5. Choice process of value chains alternatives.

A Product/Process/Resource model is the base of modelling. The product is obtained by the use of a value chain that makes the industrial potential work in order to respond to the expected functionalities (Figure 6).

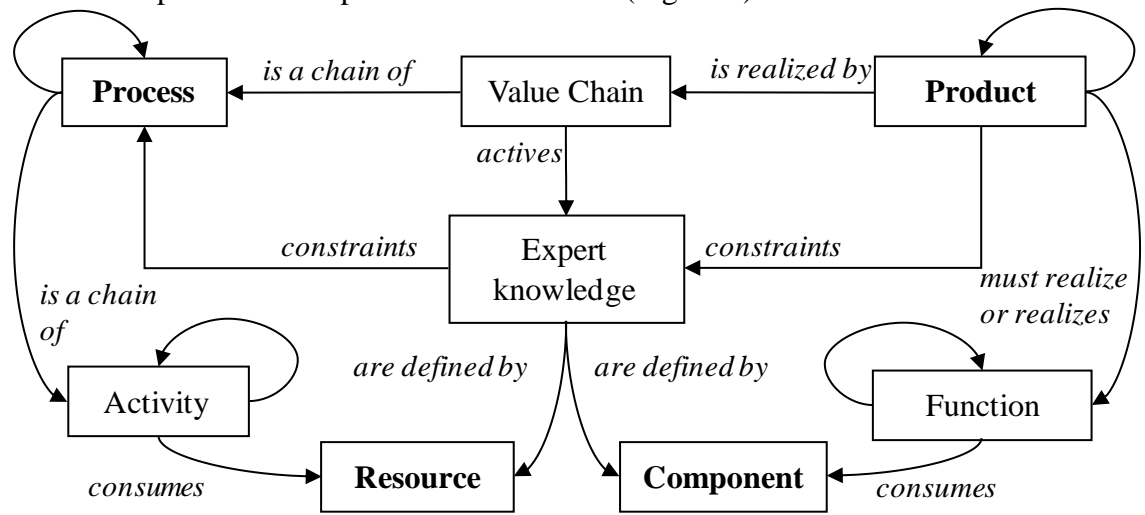

Figure 6. Structuring of the concepts for industrial system modelling.

The proposed solution that adopts value chain modelling enables the integration of all these concepts.

The effectiveness of the industrial system gives indications of the resources needed and consumed in order to give performance results; the efficiency of the industrial system is measured by the ratio between the actual performance criterion and the estimated performance criterion. In this way, the activities (that consume resources) must be studied to evaluate the global resources consumption.

**Structuring of the proposal**

The proposal is based onthe adoption of a value chain simulation approach used to face a multi-criteria analysis for decision-making support in the design and manufacturing plan phase.
The means aimed at helping the expert are divided into two groups:
- a method based on a value-chain approach that has four main steps:
    1. **formalisation** of the value-added activities and value-chain solutions responding to required product function;
    2. value chain **modelling**;
    3. **simulation** that supports the evaluation of the various value chains and that gives the value of criteria;
    4. **analysis** using the AHP process to draw a multi-criteria analysis including the judgment of the benefiting entities.
- an integrated tool that enables: the modelling of the activities in a generic manner and there chaining to specify the product realization; the simulation process of each of these value chains in order to evaluate performance criteria. This tool also provides help in analysing these results with an application tool using procedural calculations based on the value of the decision criteria and taking into account the weight attached to these criteria by the benefiting entities.

A database is required to manage and capitalise the enterprise information.

**Formalism**

The activity concept is the central concept that creates value by resource consumption. The organisation is broken down into various activities.
Activity-Based Costing method is based on the principle of causality logic and is used to improve the analysis and evaluation of the activity cost. The activity is seen as the basic component of the value chain of the enterprise [1].
Function/Behaviour/Structure model is used in the design phase for behavioural performance evaluation of product maintainability [9].
The contiguous concepts to the performance analysis are cost, quality, time and value. The concepts that support the modelling for the evaluation are: the benefiting entities, the enterprise object that are the resources, the activities, the functions, the technical solutions, the products, the value chains, the processes and the industrial potential.

**Deployment**

The deployment of such a tool is supported by a platform application based on a database, a discrete-event simulation tool and an exploitation tool.

*User interface*

The database user interface is shown in Figure 7.

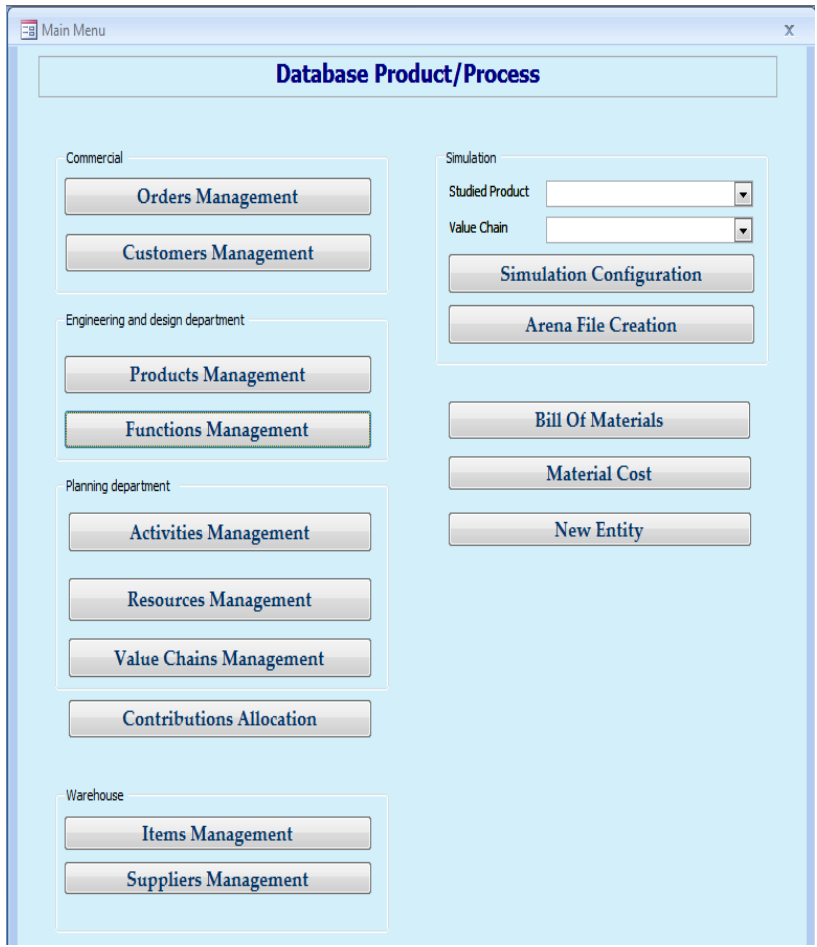

Figure 7. VCS database user interface

Experts have to record all the information relating to the order, the definition of the product and the definition of the potential value chains. This database is the basis for the creation of the activity modules needed in the simulation tool. The described value chain structures the activity sequence.

The simulation provides the performance results of the expected (ideal) process chain that are considered for the manufacture of a product. This simulation can be used by incorporating several points of view. Some of these views are implemented to respond to specific user needs relative to process choice: what activities, what sequence of activities, what resources, etc.

*Simulation module*

In order to obtain an evaluation of the performance indicators, manufacturing and support activities chaining are defined according to specific decision criteria. A flow simulation tool is then used to estimate the performance criteria of the processes [24]. In this way, comparative studies can be conducted to evaluate distinct value chains that are capable of producing the same items or complex items (interface Figure 8).

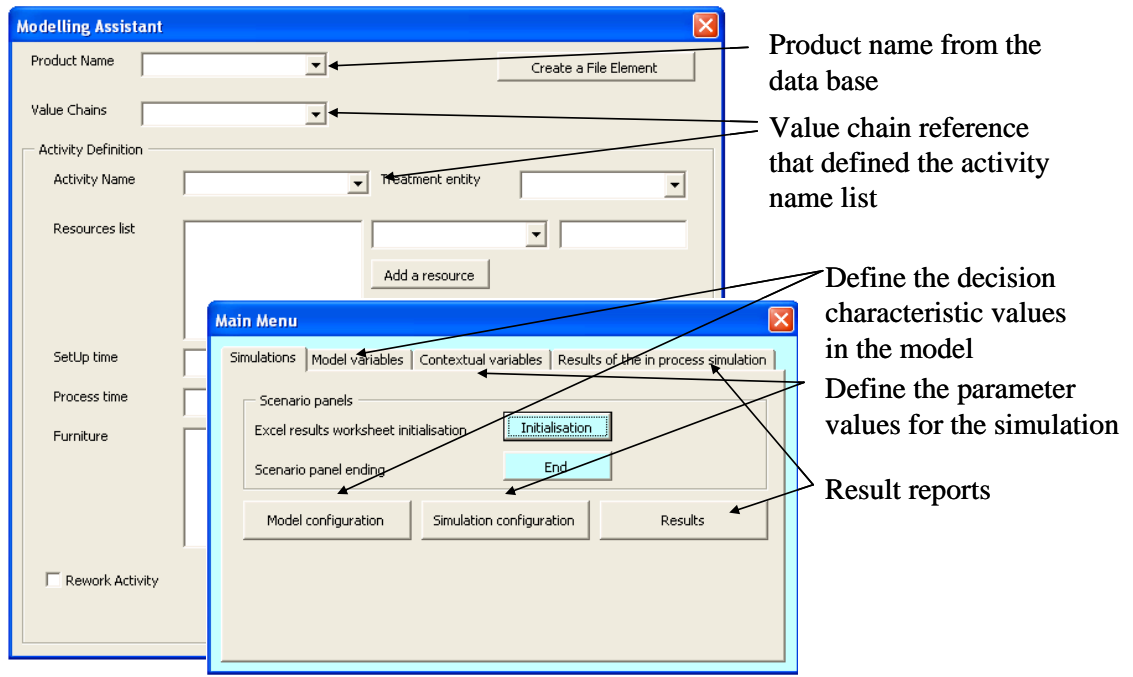

Figure 8. Simulation user interface (developed in Arena 9.0).

The application used to implement our modelling is a discrete-event flow simulation software, Arena 9.0©. The main notion is based on the entity concept. The entity is what follows the flow and what is transformed along this flow [24].

The modelling of the activity concept calls upon modelling blocks such as Submodel, Process, Decide, Hold, Assign, ReadWrite etc.

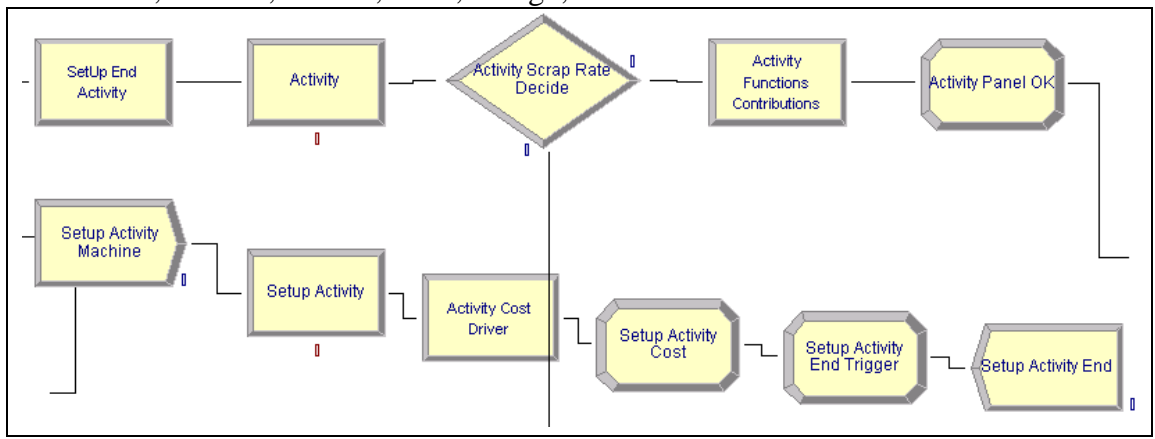

Figure 9. Arena 9.0 modules to model the activity construct.

Figure 9 is an illustration of the model implemented in Arena 9.0© to simulate an Activity. The model represents the basic element called "activity". Each activity is modelled by a similar sequence of blocks that models the resource consumption law and the performance estimate.

The data are extracted from a Microsoft Access© database in which all the activities of the enterprise are depicted by a certain number of attributes. The database also contains value chain definitions used for specific products (for instance, ballast for automotive lighting systems) and the amount of consumption of the activities cost drivers concerned by the value chain. An interface called Modelling Assistant has been developed to facilitate the modelling of the value chain in the software (Figure 8).

In this way, comparative studies (scenarios) are carried out to evaluate distinct value chains that lead to the creation of a same product or system. A useful interface supplies the launching of a series of simulations on the same product with different models or contextual variables in order to generate an Excel file that structures the comparison of these scenarios.

The simulation provides the performance results of the expected (ideal) process chain which are considered for the product manufacture. Each activity is modelled by the same model in the Arena 9.0© software (Figure 9).

This simulation can be useful for several users. For example, the information needed by the marketing manager is relevant to global indicators such as manufacturing product cost and delay (delivery time); the designer is much more interested in functions or technical solution costs; the manufacturing expert looks at activity cost; the cost analyst must obtain analytical information. Hence, some of these points of view have been implemented to respond to specific user needs.

In order to apply and validate the approach, a case study is performed in microelectronics to evaluate various possible manufacturing processes.

*AHP module*

Rabelo et al. couple the simulation system and the AHP approach to make relevant decisions on machine tool selection [19]. This demonstrates that a decision process can take advantage of simulation results.

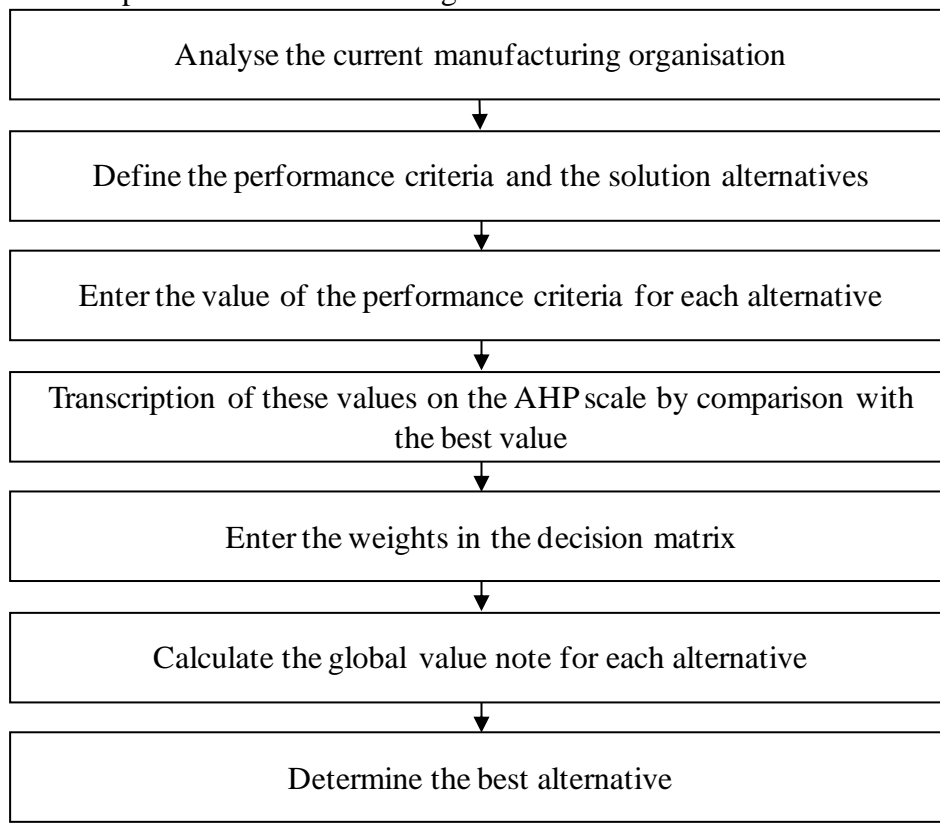

Figure 10. AHP evaluation process.

The steps of the AHP method developed are as follows (Figure 10):

*Step 1:* Structure the hierarchy (the tree) from the value components (cost, quality, time and function satisfaction) which reflect the objectives of the decision-maker viewpoint through the intermediate levels that characterize the attributes of the value component;

*Step 2:* Construct the matrix in which the results of each simulation are entered relatively to the criteria given in the lower level;

*Step 3:* Enter the results for each alternative;

*Step 4:* Construct the set of judgment matrices of each intermediate level; enter the judgments for one point of view;
*Step 5:* Construct the set of pair-wise comparison matrices for each of the lower levels (leaves of one branch);
*Step 6:* Analyse the AHP process result for each point of view.
An interface helps the user to score his preferences on performance criteria.
The results are presented in a table on which the expert can base his argumentation.

*Evaluation models*

A final notation is defined to give an assessment of the product value relative to the performance criteria: cost, risk, product conformity, time, function satisfaction.
The evaluation function of performance indicator defines the relation between the objective (o) and the measure (m) [1]. The authors address the declaration of the objective, its representation and its measure for performance analysis.
In our work, the comparison is relative. The objective is the value of the best criterion and then the evaluation function is based on the AHP scale.
For this analysis an activity-based approach is proposed. The goal is to report financial and non- financial information such as those put forward by Gunasekaran et al. [25]. In their development, Activity-Based Costing provides the structuring support (the activity) to keep non-financial information such as defect rates (quality), throughput rates (effectiveness of the industrial process) and delivery time (delay/effectiveness).

1. Cost model

One of the main elements of the modelling is the one that enables the application of the Activity-Based Costing method. This is based on the principle that activities consume resources that drive costs (Figure 11).

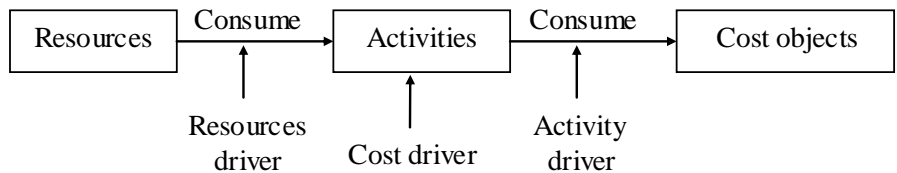

Figure 11. Process of resource consumption by cost object, according to ABC method from [26].

Consequently, a value chain has to support the modelling of resources, activities, cost objects (product and process) and three kinds of drivers, and to model the relation between them.
A cost driver is defined for each activity and thus the activity cost is expressed as the product of the consumption of this cost driver by the associated rate. The correlation method is used to determine the relevant cost driver of an activity [26]. Therefore, historical data must be exploited.
The principles of resource consumption and cost drivers come from the ABC method and the Cost Entity approach. Product cost drivers are extracted and analyzed by screening the value drivers of each activity. In order to obtain performance indicators, manufacturing activities are defined by decision criteria.
Mévellec advocates the use of non-conventional cost models for industrial contexts based on the activity principle ABC Activity-Based Costing [27].
$CI_i$: cost of the product i

$Mp_i$ : cost of the material and components incorporated in the product i
$MOD_i$ : cost of the direct labour consumed by the product i
$CI_j$ : cost of the driver of the analysis unit j (activity j)
$NbCI_{ij}$ : consumption of the cost driver j by the product i

Thus,

$$CI_i = Mp_i + MOD_i + \sum_{j=1}^{n} \left( NbCI_{ij} \times CI_j \right) \qquad (1)$$

Prerequisites:
- Activity analysis with relative expenses for all the products manufactured during a fixed period
- Cost drivers identification by activity
- Establishment of the cost of the driver for a specific activity

2. Function satisfaction

This criterion is relevant to the designer that establishes its assessment for each function and each solution. It reflects the satisfaction of the solution to respond to a specific functional requirement.

3. Delay model

The time evaluation is based on historical data since the calculation is time-rate based. For each activity, the process time to transform one entity that passes through the activity is estimated. Also, the time to process one entity is known immediately.
The time cycle is calculated by the simulation tool. It gives an average time cycle for each product.
For product value evaluation, the time criterion is defined by two measurements: cycle time per product and lead time.

4. Quality evaluation

Quality is defined by the level of the global defective rate that is represented by the number of defective products. Statistical defective rates are defined for quality controls.
A quality criterion is thus estimated by the cost of defective products.

5. Value evaluation

In the design step and during Value Analysis, the expert works on product functions. These required functions are the mediator for the choice of technical solutions and their evaluation is needed to define the final product design.
A raw interpretation of the norm of a product value definition is expressed as [28]:

$$\text{Value of a Product} = \frac{\text{User satisfaction towards the product}}{\text{Product cost}} \qquad (2)$$

This definition does not explicitly give the measurement process. For that reason, Yannou breaks down the value as the sum of the value of each function required by a product. The logical equation used to evaluate a function value is:

$$\text{Function Value} = \frac{\text{Function contribution to user satisfaction towards the product}}{\text{Cost of the solution to produce the function}} \quad (3)$$

For a function, the formula proposed by Yannou to define the function value for a specific technical solution is:

$I_j$: Importance of the function$_j$ for the product,
$S_{ji}$: Satisfaction of the solution$_j$ to respond to the function$_i$,
$C_j$: Cost of the solution$_j$.
Function contribution to user satisfaction = $I_j \times S_{ji}$

$$\text{Value of a Function}_i \text{ for a Solution}_j = \frac{I_j \times S_{ji}}{C_j} \quad (4)$$

with, $I_j$: Importance of the function$_j$ for the product, $S_{ji}$: Satisfaction of the solution$_j$ to respond to the function$_i$, $C_j$: Cost of the solution$_j$.

In that case, solution costs are known; one of our goals is to estimate the solution cost to obtain more accurate evaluation results.

The attributes of Value are Cost, Quality, Time and Function Satisfaction (see Figure 12).

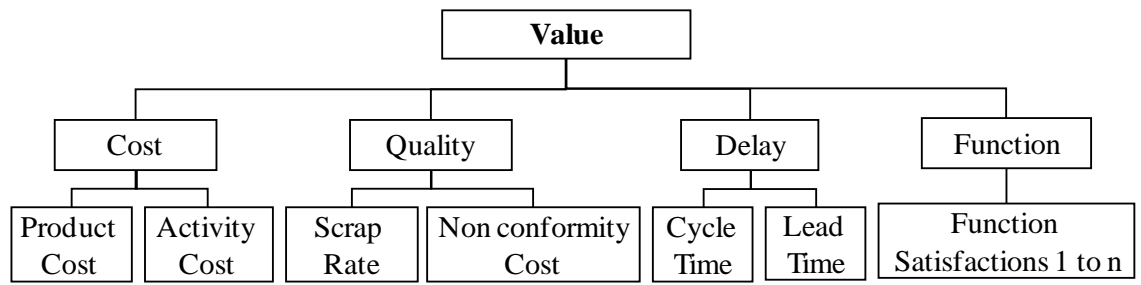

Figure 12. AHP process: Value hierarchy

So Value is measured by a weighted method, the AHP process. Benefiting entities express their point of view of the importance of a criterion for the evaluation of the product value.

# Case study

In this illustrative study, the industrial context is reduced to a manufacturing firm in microelectronics field. The set of activities (manufacturing, commercial, administrative, supplying) needs to be taken into account to simulate the global product value chain. These activities are not directly linked with the technical solutions of the product but it is necessary to achieve them in order to provide the customer with the product in best conditions. The microelectronic field studied for this value evaluation application is justified by the fact that many details and information on its manufacturing activities were available. The aim of this analysis is to check if our model would be able to support the analysis of decisions made during the design phase on manufacturing processes and to show their impacts on product value.

### First step: management and capitalisation of the enterprise information in a specific database;

The analyzed product is a ballast used in automobile lighting system for power regulation in an electrical circuit. The major functions are linked with the electrical power, should be contained in the lighting system, in order to be connect to the lighting system and to be adaptable to the lighting system technology. These functions limit the choice of technical solutions, microelectronic components and printed board.

To build the cost model, the resource consumption process has been done to define the factors that drive activity cost: time, number of pins per component, number of components per system, batch size, etc.

### Second step: modelling of the activities in a generic manner and there chaining to specify the product realization;

Globally the manufacturing process type is the assembly of microelectronic devices on a board where activities' chaining is linear except for the rework process.

### Third step: simulation process of each of these value chains in order to evaluate performance criteria by a discrete event simulation system;

Case study includes a value analysis based on two action variables: the panel size and the scrap rate of in-circuit test activity with an aim to determine which arrangement of action variables give the best result for the enterprise activity. For the experiment, two factors are taken into account as decision criteria for the design and manufacturing planning steps: first the number of boards per panel (batch factor) and second the in-circuit test scrap rate. For each of them, two potential values are acceptable and finally, four alternatives are compared.

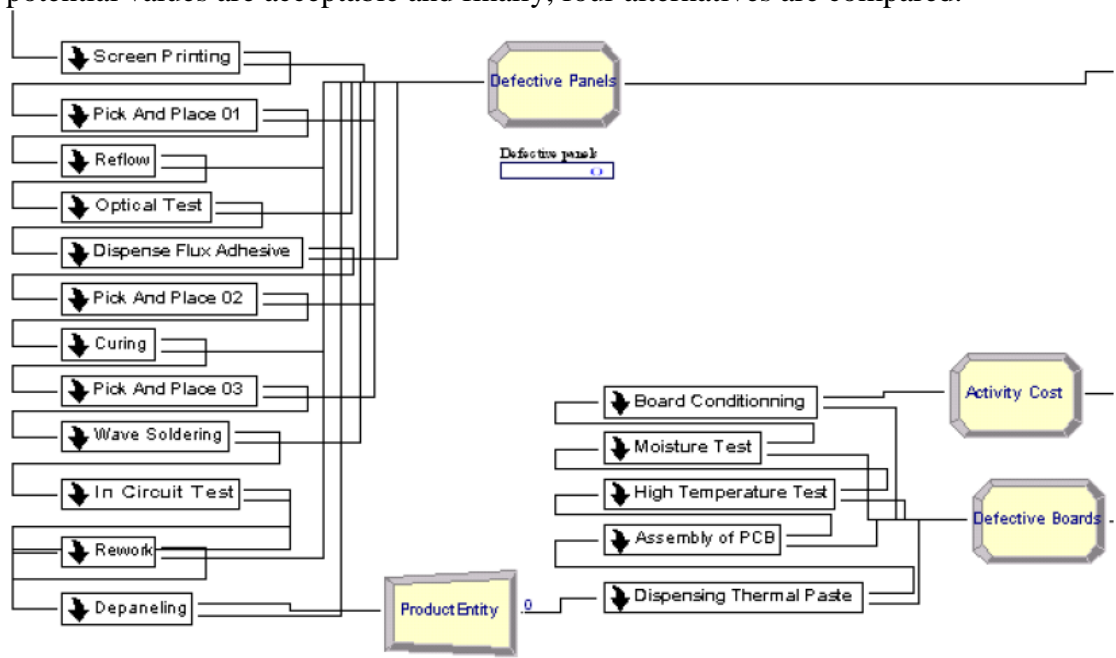

Figure 13. The simulation modelling for the evaluation of the ballast is based on a sequence of activities.

Figure 13 represents a part of the value chain modelling in Arena for the simulation of the ballast.

| Performance indicators | Scenario 1 | Scenario 2 | Scenario 3 | Scenario 4 |
|---|---|---|---|---|
| Total Activity Cost (€) | 165 080,17 | 128 315,78 | 96 735,12 | 79 149,88 |
| Unit Material Cost (€) | 9,41 | 9,41 | 9,24 | 9,24 |
| Material Cost (€) | 78 121,82 | 78 018,31 | 74 834,76 | 74 890,20 |
| Total Cost (€) | 243 201,99 | 206 334,09 | 171 569,88 | 154 040,08 |
| Unit Cost (€) | 29,29 | 24,89 | 21,18 | 19,01 |
| Total Cycle Time (s) | 198 868 | 194 117 | 219 684 | 212 287 |
| Number of Defective Boards | 423 | 612 | 180 | 156 |
| Lead Time (s) | 498,7 | 498,7 | 261,9 | 251,6 |
| Delay Time (s) | -2 732 (in advance) | -7 483 (in advance) | 18 084 (delay) | 10 687 (delay) |

Table 2 – Results given by the simulation of the product value chains

In Table 2, results directly provided by the simulation module are tabulated under four scenarios. Notice that the delay time is evaluated relatively to the delivery time imposed by the client.

**Fourth step: analysis using the AHP process to include the benefiting entities judgment.**

A decision must be taken according to these types of results but decision is not obvious; a rough comparison of the results by performance criteria (cost, cycle time, quality indicator) gives a basic classification in Table 3.

Classification of the scenarios according to

| Performance indicators | Scenario 1 | Scenario 2 | Scenario 3 | Scenario 4 |
|---|---|---|---|---|
| Unit Cost | 4 | 3 | 2 | 1 |
| Total Cycle Time | 2 | 1 | 4 | 3 |
| Number of Defective Boards | 3 | 4 | 2 | 1 |

Table 3 – Basic classification

Due to various positions of the scenarios according to the performance indicator chosen, the decision must be a compromise. This compromise can be finding by integrating the benefiting entities point of view through AHP.

The AHP process treats the simulation results presents in Table 3 to establish the results of the Table 4.

| Scenario | Designer | Manufacturer | Client | Shareholders |
|---|---|---|---|---|
| Scenario 1 | **0,3912** | **0,3912** | 0,3794 | 0,1941 |
| Scenario 2 | 0,3894 | 0,3719 | **0,3854** | 0,2405 |
| Scenario 3 | 0,1015 | 0,1070 | 0,0981 | 0,2106 |
| Scenario 4 | 0,1174 | 0,1298 | 0,1371 | **0,3549** |

Table 4 - AHP analysis results for each benefiting entity.

These last results demonstrate that:
    1) scenario 1 is the best regarding preferences of designer and manufacturer;

2) scenario 2 is the best regarding preferences of client and scenario 1 is at the second place;
3) from the shareholders point of view, scenario 4 is the best since cost is the primordial criterion.

Globally, scenario 1 is the one to be chosen in order to best satisfy all the benefiting entities. The results of scenario 1 can be then analysed to ensure that unit cost is higher but the client shall be served in time.

The purpose of this case analysis is to provide comparison indicators to take a decision on a dimensional parameter and on the choice of a manufacturing resource. This kind of approach can be applied to decision support on technical choices. From the same functional definition, depending to the enterprise technologies possibilities and limits, the optimal product-process solution can be evaluated, taking into account all the benefiting entities point of view. Technical solutions and choices can be then adjusted to strategically axes.

## Conclusion

The improvement expected is to enhance the performance evaluation process of a product and its manufacturing process. A value-based approach is used to define manufacturing process chain modelling and is applied to support its performance evaluation. The evaluation methodology is assisted by our specific simulation tool proposal: VCS. This tool and methodology is based on a modelling of the manufacturing enterprise system through the concepts and their relations. This entity relation model structures the information system.

Value chain modelling gives an appropriate support to performance evaluation. This kind of modelling is understood by all the actors. The basic principle of activity links (or is linked to) all the modelling elements.

Moreover, each participant, in the decision process, can express his/her point of view which is taken into account in the analysis process. The first experiment shows that the use of VCS provides consistent results.

Our final purpose would be to simulate value chains to generate alternative chains coupled with optimization methods to find the best solutions. We limit our work to measurable values and to manufacturing steps. Then other aspects could be investigated to consider less observable information in order to tackle the whole product life cycle and non-tangible value aspects. A new study covering logistic processes and/or support activities must be carried out to assess the relevance of our approach.

Moreover, complementary research is being done to integrate risk factors in our model. As a perspective, integration of resource competencies can be done in order to measure the impact of resource choice in the value system [10]. Also, others value metrics can be addressed [29].